\documentstyle[12pt]{article}
\newcommand{\nonum}{\nonumber}
\newcommand{\nn}{\nonumber \\}
\renewcommand{\o}{\over}
\newcommand{\ri}{\right}
\newcommand{\lf}{\left}
\newcommand{\Bl}{\biggl}
\newcommand{\Br}{\biggr}
\newcommand{\bl}{\Bigl}
\newcommand{\br}{\Bigr}
\newcommand{\T}{{\rm T}}
\newcommand{\Ts}{{\rm T}^{\ast}}

\newcommand{\del}{\partial}
\newcommand{\cob}{\delta}    
\newcommand{\al}{\alpha}

\newcommand{\bt}{\beta}

\newcommand{\ep}{\epsilon}
\newcommand{\vep}{\varepsilon}
\newcommand{\th}{\theta}
\newcommand{\Ga}{\Gamma}

\newcommand{\si}{\sigma}
\newcommand{\ka}{\kappa}

\newcommand{\la}{\lambda}

\newcommand{\riya}{\rightarrow}

\newcommand{\tr}{{\rm{tr}}}

\newcommand{\ps}{\not\!\!P}
\newcommand{\ket}{\langle}
\newcommand{\bra}{\rangle}

\newcommand{\dph}{\phi^{\dagger}}
\renewcommand{\^}{\hat}
\newcommand{\half}{{1 \over 2}}
\renewcommand{\b}[1]{\bar{#1}}
\renewcommand{\t}[1]{\tilde{#1}}
\newcommand{\rt}{\sqrt{2}}

\begin{document}
\vskip 0.5 truecm
{\baselineskip=14pt
 \rightline{ \vbox{
  \hbox{UT-785}
       }}}

\vfill
\begin{center}
{\Large\bf BRST gauge fixing and\\ the algebra of  global supersymmetry }
\end{center}
\vskip 1 truecm
\centerline{ Kazuo Fujikawa and Kazumi Okuyama}
\vskip .5 truecm
\centerline {\it Department of Physics,University of Tokyo}
\centerline {\it Bunkyo-ku,Tokyo 113,Japan}
\vskip 0.5 truecm
\centerline {\sf fujikawa@hep-th.phys.s.u-tokyo.ac.jp
~,~ okuyama@hep-th.phys.s.u-tokyo.ac.jp}

\makeatletter
\@addtoreset{equation}{section}
\def\theequation{\thesection.\arabic{equation}}
\makeatother

\vfill

\begin{abstract}
A global supersymmetry (SUSY) in  supersymmetric gauge theory is generally
broken by gauge fixing. A prescription to extract physical information
from such  SUSY algebra broken by gauge fixing is analyzed  in path
integral framework. If $\delta_{SUSY}\delta_{BRST}\Psi =
\delta_{BRST}\delta_{SUSY}\Psi$ for a gauge fixing ``fermion'' $\Psi$, the 
SUSY charge density is written as a sum of the piece which is naively
expected without gauge fixing and a BRST exact piece. If
$\delta_{SUSY}\delta_{SUSY}\delta_{BRST}\Psi =
\delta_{BRST}\delta_{SUSY}\delta_{SUSY}\Psi$, the equal-time 
anti-commutator of SUSY charge is written as a sum of a physical piece and
a BRST exact piece. We illustrate these properties for $N=1$ and $N=2$ 
supersymmetric Yang-Mills theories and for a $D=10$ massive superparticle 
(or ``D-particle'') where
the $\kappa$-symmetry provides extra complications.

\par
\end{abstract}
\vskip 2cm

\newpage 
\baselineskip 7mm
\section{Introduction}

In some of the applications of supersymmetric (SUSY) gauge theory
\cite{wess}, 
the SUSY algebra, in particular, its possible central extension provides 
important
physical information\cite{witten-olive}. On the other hand, the global SUSY
in generic supersymmetric 
gauge theory is generally broken by gauge fixing, at least in the component
formulation in the Wess-Zumino gauge.  The SUSY charge thus
generally becomes {\em time-dependent} after gauge fixing.  In such a
situation, 
it is important to specify how to extract physical information from the
{\em broken} charge 
algebra. In the present note, we present a general prescription how to
extract
physical information from the {\em equal-time} anti-commutator of such
broken 
SUSY charge density in the framework of path integral and the
Bjorken-Johnson-Low (BJL) prescription \cite{bjl}. 
The quantities in path integral ,
which are defined in 
terms of $\Ts$-product , are converted into those in the conventional
$\T$-product by BJL prescription, which in turn readily provide information
about the equal-time (anti-) commutator. 

The essence of our analysis  is summarized as follows:  
If one defines the gauge fixing Lagrangian by ${\cal L}_{g} = \delta_{BRST}
\Psi$, where $\Psi$ is a gauge fixing ``fermionic'' function, the SUSY
charge density $J^{0}_{SUSY}$ is written as 
\begin{equation}
J^{0}_{SUSY} = J^{0}_{SUSY(naive)} +\  BRST\  exact\  piece
\label{mod-charge}
\end{equation}
if 
\begin{equation}
\delta_{SUSY}\delta_{BRST}\Psi = \delta_{BRST}\delta_{SUSY}\Psi
\label{SB=BS}
\end{equation}
where $\delta_{SUSY}$ is a localized (coordinate dependent) SUSY 
transformation, and  $J^{0}_{SUSY(naive)}$ is the naive charge density
expected before gauge fixing. 

If one has the relation with another SUSY transformation
\begin{equation}
\delta^{\prime}_{SUSY}\delta_{SUSY}\delta_{BRST}\Psi =
\delta_{BRST}\delta^{\prime}_{SUSY}\delta_{SUSY}\Psi
\label{SSB=BSS}
\end{equation}
one obtains the equal-time anti-commutation relation
\begin{equation}
\{ Q_{\alpha}(x^{0}), Q_{\beta}(y^{0})\}_{x^{0} = y^{0}} = physical\  piece
+ BRST\ exact\ piece
\label{uptoB-comm}
\end{equation}
where the  physical piece  as well as the BRST exact piece are obtained
from  a SUSY transformation of the SUSY
charge density in path integral formulation. 
These properties hold when one assigns all the unphysical  (ghost) 
particles 
to be SUSY {\em scalar}, namely, they are not transformed under SUSY 
transformation.

An advantage of  this path 
integral formulation is that one can readily identify the possible presence
of generic ``Schwinger terms'' such as  the physical central
extension. It turns out that the
second property in (\ref{uptoB-comm}) holds for a more general
 class of theories and
gauge fixing than the 
first property in (\ref{mod-charge}). The properties (\ref{mod-charge}) and
(\ref{SB=BS}) have been analyzed
in the past in Refs.\cite{dewit} and \cite{holten}. To our knowledge, 
a systematic analysis of 
(\ref{SSB=BSS}) and (\ref{uptoB-comm}) has not been performed before.

In the following, we illustrate these properties for $N=1$ and $N=2$
supersymmetric
Yang-Mills theories  and for a $D=10$ massive  superparticle which may
 be regarded as a ``D-particle''.
In the latter case, the so-called $\kappa$- symmetry gives rise to
additional technical complications.

\section{$N=1$ super Yang-Mills theory}

As a first  example, we consider  supersymmetric  Yang-Mills theory . 
The Lagrangian is given by 
\begin{equation}
{\cal L}_0 = -{1\o g^2}2\tr\lf({1\over 4}F_{mn}F^{mn}+i\bar{\la}\bar{\si}^m
D_m\la\ri)
\label{n=1Lag}
\end{equation}
where
\begin{eqnarray}
&&D_m\la=\del_m\la-i[A_m,\la] \nonum \\
&&F_{mn}=\del_mA_n-\del_nA_m-i[A_m,A_n]
\end{eqnarray}
Here we normalized the generators $T^{a}$ of gauge symmetry by $\tr
T^{a}T^{b} =\frac{1}{2}\delta^{ab}$.
The Landau-type  gauge
, for example,  is defined by 
\begin{equation}
{\cal L}_g = 2\tr(B\del_mA^m-i\bar{c}\del_mD^mc)=2\tr\cob_{BRST}
(\bar{c}\del_mA^m)
\label{L-g}
\end{equation}
The SUSY transformation rules are 
\begin{eqnarray}
&&\cob_{SUSY} A_m=i(\bar{\xi}\bar{\si}_m\la-\bar{\la}\bar{\si}_m\xi)
\nonum \\
&&\cob_{SUSY} \la = \si^{mn}\xi F_{mn}\nonum \\
&&\cob_{SUSY} \bar{\la}=-\bar{\xi}\bar{\si}^{mn}F_{mn}
\end{eqnarray}
and the BRST transformation rules are defined by using a Grassmann 
parameter $\ep$ as 
\begin{eqnarray}
&&\cob A_m=i\ep D_mc=i\ep(\del_m c-i[A_m,c]) \nonum \\
&&\cob \la=-[\ep c,\la], \ \ \  \cob \bar{\la}=-[\ep c, \bar{\la}] \nonum
\\
&&\cob c =-{1\over 2}[\ep c,c] \nonum \\
&&\cob \bar{c}=\ep B, \ \ \  \cob B=0 
\end{eqnarray}
Here $c$ and $\bar{c}$ are the Faddeev-Popov ghost and anti-ghost, and $B$
is the Nakanishi-Lautrup field. 
We can confirm that $\cob_{BRST}$ and $\cob_{SUSY}$ commute on $A_{\mu}$
and $\la$ by
the above transformation rules
\begin{eqnarray}
&&\cob_{BRST} \cob_{SUSY} A_{\mu}= \cob_{SUSY}\cob_{BRST} A_{\mu}=-[\ep c,
\cob_{SUSY} A_{\mu}] \nonum\\
&&\cob_{BRST} \cob_{SUSY} \la= \cob_{SUSY}\cob_{BRST}\la=-[\ep
c,\cob_{SUSY}\la]
\label{sb-commute}
\end{eqnarray}
In other words, the gauge fixing fermion $\Psi = \bar{c}\partial_{m}A^{m}$
in (\ref{L-g}) satisfies
\begin{eqnarray}
\delta_{BRST}\delta_{SUSY}\Psi& =&\delta_{BRST}
\bar{c}\partial_{m}(\delta_{SUSY}A^{m})\nonumber\\
&=& B\partial_{m}(\delta_{SUSY}A^{m}) +
\bar{c}\partial_{m}(\delta_{BRST}\delta_{SUSY}A^{m})\nonumber\\
&=& B\partial_{m}(\delta_{SUSY}A^{m}) +
\bar{c}\partial_{m}(\delta_{SUSY}\delta_{BRST}A^{m})\nonumber\\ 
&=& \delta_{SUSY}(\delta_{BRST}(\bar{c}\partial_{m}A^{m}))\nonumber\\
&=& \delta_{SUSY}\delta_{BRST}\Psi
\end{eqnarray}
and similarly
\begin{equation}
\delta_{BRST}\delta^{\prime}_{SUSY}\delta_{SUSY}\Psi =
\delta^{\prime}_{SUSY}\delta_{SUSY}\delta_{BRST}\Psi
\end{equation}
even for the localized (coordinate dependent) SUSY transformation. In fact 
these properties hold for a more general class of gauge fixing fermion
$\Psi
= \bar{c}F( A_{m}, \lambda)$. Note that all the un-physical particles are
assigned to be SUSY scalar, namely, they are not transformed under SUSY
transformation. These properties suggest that 
  the supercurrent and superalgebra have the structure (\ref{mod-charge})
and (\ref{uptoB-comm}), which can be confirmed below.

The variation of the total Lagrangian ${\cal L} = {\cal L}_{0} + {\cal
L}_{g}$
under SUSY transformation with 
position dependent parameters is
\begin{equation}
\cob_{SUSY} {\cal L}=\del^m \bar{\xi}\bar{S}_m+\bar{\xi}\bar{J}+
S_m  \del^m\xi +J\xi
\label{s-trf-YM}
\end{equation}
where the SUSY currents and SUSY violating source terms are given by 
\begin{eqnarray}
&&S_m=-{2i\o g^2}\tr(\bar{\la}\bar{\si}_m\si_{kl}F^{kl})
-2i\tr\cob_{BRST}(\bar{c}\bar{\la}\bar{\si}_m)
=-{2i\o g^2}\tr(2\bar{\la}\bar{\si}^nF^+_{nm})
-2i\tr\cob_{BRST}(\bar{c}\bar{\la}\bar{\si}_m)
\nonum \\
&&\bar{S}_m=-{2i\o g^2}\tr(F^{kl}\bar{\si}_{kl}\bar{\si}_m\la)
 -2i\tr\cob_{BRST} (\bar{c}\bar{\si}_m\la)
=-{2i\o g^2}\tr(2 F^-_{mn}\bar{\si}^n\la)
-2i\tr\cob_{BRST}(\bar{c}\bar{\si}_m\la)
\nonum \\
&&J=-2i\tr\cob_{BRST} (\bar{c}\del^m\bar{\la}\bar{\si}_m) \nonum \\
&&\bar{J}=2i\tr\cob_{BRST} (\bar{c}\bar{\si}_m\del^m\la)
\label{sym-current}
\end{eqnarray}
These SUSY currents $S_m$ and $\bar{S}_m$ have the structure indicated in
(\ref{mod-charge}). Here we defined  
\begin{eqnarray}
&&F^{\pm}_{mn}=\half(F_{mn}\pm \^{F}_{mn})=
\half\lf(F_{mn}\pm i\half\vep_{mnkl}F^{kl}\ri)
\end{eqnarray}
where $\^{F}=\hat{\ast}F=i\ast \!F$. In Lorentz metric
$\hat{\ast}^2F=F$, and $F^{\pm}$ are a self-dual (or  anti-self-dual) part
of $F$ ,i.e., $\hat{\ast}F^{\pm}=\pm F^{\pm}$.

A Ward-Takahashi identity  for supersymmetry  is obtained by
starting with
\begin{equation}
\langle S_{a}(y)\eta\rangle \equiv \int d\mu S_{a}(y)\eta e^{i\int {\cal
L}d^{4}x}
\end{equation}
with a global Grassmann parameter $\eta$ and performing the change of field
variables under the localized SUSY transformation parametrized by $\xi (x)$
and $\bar{\xi} (x)$.  By taking the variational derivative with respect to
$\xi (x)$
and $\bar{\xi}(x)$ , one obtains W-T identities. For notational simplicity,
we here write the identities obtained by multiplying {\em global}
$\xi$ and $\bar{\xi}$ anew:  
\begin{eqnarray} 
&&\int d^3 x \xi [\del_0\langle \Ts S^0(x)S_a(y)\eta \rangle -
\langle \Ts J(x)S_a(y)\eta\rangle ]\nonumber\\
&+& \int d^3 x \bar{\xi} [\del_0 \langle \Ts \bar{S}^0(x)
S_a(y)\eta\rangle - \langle \Ts \bar{J}(x) S_a(y)\eta \rangle] \nonum
\\
&=& \cob(x^0-y^0)\lf\ket 2\bar{\xi}\bar{\si}^m\eta\bl(\tilde{T}_{ma}(y)
-2\tr\cob_{BRST}(\bar{c}F^-_{ma})(y)\br)+\xi M_a(y)\eta \ri\bra \nonum \\
&&-{\del\over \del x^0}\cob(x^0-y^0)
\lf\ket 4\tr\bl(\bar{\la}\bar{\si}_a\si^{0l}\eta\cob_{SUSY} A_l\br)(y)
\ri\bra
\label{Ts-YM}
\end{eqnarray}
where we defined
\begin{equation} 
\xi M_a\eta = (i \xi\si_{an}\eta  - \frac{3}{2}i \eta_{an}\xi\eta )2\tr [ 
\bar{\lambda}\bar{\sigma}^{n}\sigma^{k}D_{k}\bar{\lambda}]
\label{center-YM}
\end{equation}
and $\tilde{T}_{ma}$ is the energy-momentum tensor for supersymmetric 
Yang-Mills theory
\begin{eqnarray} 
&&\tilde{T}_{ma}=T^{YM}_{ma}+ T^{\la}_{ma} \nonum \\
&&g^2T^{YM}_{ma}=2\tr\lf(F_{mk}F_a^{~k}-{1\over 4}\eta_{ma}F_{kl}F^{kl}\ri)
\nonum \\
&& g^2T^{\la}_{ma}={i\over 4}2\tr(\bar{\la}\bar{\si}_mD_a\la+
3\bar{\la}\bar{\si}_aD_m\la)
-{1\over 4}\eta_{ma}2\tr(i\bar{\la}\bar{\si}^kD_k\la) \nonum \\
&&~~~~~~~~~~-{1\over 4}\vep_{makn}2\tr(\bar{\la}\bar{\si}^nD^k\la)
\label{EM-YM}
\end{eqnarray} 

We here recall the basic idea of the BJL prescription \cite{bjl}. The
correlation
function defined by $\Ts$-product
\begin{equation}
\langle \Ts A(x)B(y)\rangle
\end{equation}
can be replaced by the conventional $\T$-product if
\begin{equation}
\lim_{q^{0}\rightarrow \infty}\int d^{4}x e^{iq(x-y)}\langle  \Ts
A(x)B(y)\rangle = 0
\end{equation}
If the above quantity does not vanish, one defines the $\T$-product by
\begin{eqnarray}
\int d^{4}x e^{iq(x-y)}\langle  \T  A(x)B(y)\rangle &\equiv& 
\int d^{4}x e^{iq(x-y)}\langle  \Ts A(x)B(y)\rangle \nonumber\\
 &-& \lim_{q^{0}\rightarrow \infty} \int d^{4}x e^{iq(x-y)}\langle 
\Ts A(x)B(y)\rangle
\end{eqnarray}
In either case we have
\begin{equation}
\lim_{q^{0}\rightarrow \infty}\int d^{4}x e^{iq(x-y)}\langle  \T
A(x)B(y)\rangle = 0
\end{equation}
which defines the $\T$-product in general.

In the present context, the term proportional to ${\del\over \del
x^0}\cob(x^0-y^0)$ in the right-hand side is subtracted away if one uses
the $\T$-product in eq.(\ref{Ts-YM}). We then obtain
\begin{eqnarray} 
&&\int d^3 x \xi [\del_0\langle \T S^0(x)S_a(y)\eta \rangle -
\langle \T J(x)S_a(y)\eta\rangle ]\nonumber\\
&+& \int d^3 x \bar{\xi} [\del_0 \langle \T \bar{S}^0(x) S_a(y)\eta\rangle
-
\langle \T \bar{J}(x) S_a(y)\eta \rangle ]\nonum \\
&=& \cob(x^0-y^0)\lf\ket2\bar{\xi}\bar{\si}^m\eta\bl(\tilde{T}_{ma}(y)
-2\tr\cob_{BRST}(\bar{c}F^-_{ma})(y)\br)+\xi M_a(y)\eta \ri\bra 
\end{eqnarray}
If one performs the explicit time derivative operation in the left-hand
side 
and if one uses the current conservation condition following from 
 (\ref{s-trf-YM})
\begin{equation}
\int d^{3}x (\partial_{0}S^{0}(x) - J(x)) = \int d^{3}x
(\partial_{0}\bar{S}^{0}(x) - \bar{J}(x)) = 0
\end{equation}
one obtains  the following commutation relations
\begin{eqnarray}
&&\bl[\bar{\xi}\bar{Q}(x^0),S_a(y)\eta\br]_{x^0=y^0} =
2\bar{\xi}\bar{\si}^m\eta\bl(\tilde{T}_{ma}
-2\tr\cob_{BRST}(\bar{c}F^-_{ma})\br)(y)
\label{bQQ}\\
&&\bl[\xi Q(x^0),S_a(y)\eta\br]_{x^0=y^0} =\xi M_a(y) \eta 
\label{QQ}
\end{eqnarray} 
Eq.(\ref{bQQ}) shows that SUSY algebra at equal-time closes up to a BRST
exact term.
If the right-hand side of  Eq.(\ref{QQ}) does not vanish, it would
represent a possible ``central extension of N=1 SUSY
algebra''. Note that nowhere in our calculation the
equations of motion have been used. The right-hand side of  Eq.(\ref{QQ}) 
vanishes if one uses the (safe) equation of motion in Eq.(\ref{center-YM}).
We thus recover the conventional SUSY algebra defined in Poincare invariant
vacuum
, as is described in (\ref{uptoB-comm}). A further comment on the 
central extension will be given  in Section 5.

In passing, we here note several useful relations in our path integral 
manipulation. 
To obtain (\ref{sym-current}) and also identify the energy-momentum tensor 
(\ref{EM-YM}), we use the following identities,
\begin{eqnarray}
&&\bar{\si}^l\si^m\bar{\si}^n=-\eta^{lm}\bar{\si}^n-\eta^{mn}\bar{\si}^l+
\eta^{ln}\bar{\si}^m-i\vep^{lmnk}\bar{\si}_k \nonum \\
&&F_{kl}\bar{\si}^{kl}\bar{\si}^m=2F_{-}^{mn}\bar{\si}_n \nonum \\
&&\bar{\si}^m\si^{kl}F_{kl}=2\bar{\si}_nF_+^{nm}
\end{eqnarray}
For the general 2-form $A_{mn}$ and $B_{mn}$,  we have the identities
\begin{eqnarray}
&&A^+_{mn}B^{-mn}=0 \nonum \\
&&A^{\pm}_{mk}B^{\pm k}_n+B^{\pm}_{mk}A^{\pm k}_n
=\half \eta_{mn}A^{\pm}_{kl}B^{\pm kl}\nonum \\
&&A^+_{mk}B^{-k}_n=B^-_{mk}A^{+k}_n=A^-_{mk}B^{+k}_n
\end{eqnarray}
Using these identities, the energy-momentum tensor of gauge field $T^{{\rm
YM}}_{mn}$ in (\ref{EM-YM}) is  written as
\begin{equation}
T^{{\rm YM}}_{mn}={4\o g^2}\tr F^+_{mk}F_n^{-k}={4\o g^2}\tr
F^-_{mk}F_n^{+k}
\end{equation}
In the evaluation of (\ref{center-YM}), we used the relation
\begin{equation}
\xi\si^{mk}\eta\tr[\bar{\lambda}\bar{\si}_{ma}D_k\bar{\lambda}] =
 \xi\si_{ma}\eta\tr[\bar{\lambda}\bar{\si}^{mk}D_k\bar{\lambda}]
\end{equation}

\section{$N=2$ super Yang-Mills theory}

We next analyze the superalgebra of $N=2$ super Yang-Mills theory.
The Lagrangian of $N=2$ super Yang-Mills theory is given by Ref.\cite{N=2}.
\begin{eqnarray}
{\cal L}_0 &=& {2 \o g^2}\tr\Bl(-{1\o4}F^{mn}F_{mn}-D_m\dph D_m\phi
-i\b{\la}\b{\si}^mD_m\la-i\b{\psi}\b{\si}^mD_m\psi \nn
&&+i\rt (\b{\psi}[\b{\la},\phi] +[\la,\dph]\psi)-\half[\phi,\dph]^2\Br)
+{\th\o 16\pi^2}\tr F_{mn}\t{F}^{mn}
\label{n2-lag}
\end{eqnarray}
 where $\t{F}^{mn}=\half\vep^{mnkl}F_{kl}$.
This Lagrangian is invariant under $N=2$ SUSY. The first SUSY
transformation $\cob^{(1)}$ is defined by 
\begin{equation}
\begin{array}{lrl}
\cob^{(1)}_{\xi}\phi = \rt \xi\psi  &,& \cob^{(1)}_{\xi}\dph = 0 \\
\cob^{(1)}_{\xi}\psi = 0 &,& \cob^{(1)}_{\xi}\b{\psi} =-i\rt
D_m\dph\xi\si^m \\
\cob^{(1)}_{\xi}\la = (\si^{kl}F_{kl}+i[\phi,\dph])\xi &,&
\cob^{(1)}_{\xi}\b{\la} = 0 \\
\cob^{(1)}_{\xi}A_m = -i\b{\la}\b{\si}_m\xi &
\end{array}
\end{equation}
We write only the SUSY transformation with the parameter $\xi$;
SUSY with the parameter $\b{\xi}$ is given by its complex conjugation.
The variables $\la$ and $\psi$ form an $SU(2)_R$ doublet. 
The second SUSY transformation $\cob^{(2)}$ is obtained from  $\cob^{(1)}$
by 
the $SU(2)_R$ rotation  $(\la,\psi)\riya (\psi,-\la)$, 
\begin{equation}
\begin{array}{lrl}
\cob^{(2)}_{\xi}\phi = -\rt \xi\la  &,& \cob^{(2)}_{\xi}\dph = 0 \\
\cob^{(2)}_{\xi}\psi = (\si^{kl}F_{kl}+i[\phi,\dph])\xi &,&
\cob^{(2)}_{\xi}\b{\psi} = 0 \\
\cob^{(2)}_{\xi}\la = 0 &,&
\cob^{(2)}_{\xi}\b{\la} = i\rt D_m\dph\xi\si^m  \\
\cob^{(2)}_{\xi}A_m = -i\b{\la}\b{\si}_m\xi &
\end{array}
\end{equation}

The gauge fixing term for the Landau gauge is given by 
\begin{equation}
{\cal L}_g=2\tr
(B\del_mA^m-i\b{c}\del_mD^mc)=2\tr\cob_{BRST}(\b{c}\del_mA^m)
\end{equation}
BRST transformation rules are  defined by
\begin{eqnarray}
&&\cob A_m=i\ep D_mc \nn
&&\cob\la=-[\ep c,\la] ~,~\cob\psi=-[\ep c,\psi] \nn
&&\cob\phi=-[\ep c,\phi] \nn
&&\cob c=-\half[\ep c,c] \nn
&&\cob \b{c}=\ep B ~,~\cob B=0
\end{eqnarray}
We can confirm that $\cob_{BRST}$ and $\cob_{SUSY}$ thus defined commute on
$\phi,\psi,\la$ and$A_m$. See (\ref{sb-commute}). We thus conclude that 
the general analyses (\ref{mod-charge}) - (\ref{uptoB-comm}) apply 
to the present case, as is explicitly demonstrated
below.

The variation of the total Lagrangian ${\cal L}={\cal L}_0+{\cal L}_g$
under SUSY transformation with position dependent parameters $\xi$ ($A=
1,2$) is
\begin{equation}
\cob^{(A)}_{\xi}{\cal L}=S_m^{(A)}\del^m\xi+J^{(A)}\xi
\end{equation} 
where the supercurrents and SUSY violating densities are given by
\begin{eqnarray}
S_m^{(1)}\eta &=&{2i\o g^2}\tr\lf[-\b{\la}\b{\si}_m(\si^{kl}F_{kl}
+i[\phi,\dph])\eta-i\rt D_k\dph\eta\si^k\b{\si}_m\psi\ri]
        -2i\tr\cob_{BRST}(\b{c}\b{\la}\b{\si}_m\eta) \nn
  &=&{2i\o g^2}\tr \cob_{\eta}^{(1)}(-\b{\la}\b{\si}_m\la
     +\b{\psi}\b{\si}_m\psi)-2i\tr\cob_{BRST}(\b{c}\b{\la}\b{\si}_m\eta) \\
S_m^{(2)}\eta &=&{2i\o g^2}\tr\lf[-\b{\psi}\b{\si}_m(\si^{kl}F_{kl}
+i[\phi,\dph])\eta-i\rt D_k\dph\eta\si^k\b{\si}_m\la\ri] 
        -2i\tr\cob_{BRST}(\b{c}\b{\psi}\b{\si}_m\eta)\nn
  &=&{2i\o g^2}\tr \cob_{\eta}^{(2)}(-\b{\psi}\b{\si}_m\psi
      +\b{\la}\b{\si}_m\la)-2i\tr\cob_{BRST}(\b{c}\b{\psi}\b{\si}_m\eta) \\
J^{(1)}&=&-2i\tr\cob_{BRST}(\b{c}\del_m\b{\la}\b{\si}^m) \\
J^{(2)}&=&-2i\tr\cob_{BRST}(\b{c}\del_m\b{\psi}\b{\si}^m)
\end{eqnarray}
We multiplied supercurrents by a global Grassmann parameter $\eta$ to form
Lorentz scalar quantities. 

An interesting property of  $N=2$ superalgebra is  the general existence of
the
central charge, which can appear in
$\{Q_{\al}^{(1)},Q_{\bt}^{(2)}\}$\cite{wess}. To
calculate this anti-commutator, we start with the Ward-Takahashi identity
 for supersymmetry in path integral formulation, which is obtained from the
expression 
$\ket S_m^{(2)}(y)\eta\bra$ and a change of integration variables
corresponding to localized supersymmetry $\cob^{(1)}_{\xi}$,
\begin{eqnarray}
&&\int d^3x ~\xi\bl[\del_0\ket\Ts  S^{0(1)}(x)S_m^{(2)}(y)\eta\bra
-\ket\Ts  J^{(1)}(x)S_m^{(2)}(y)\eta\bra\br] \nn
&=&-2\rt i\cob(x^0-y^0)\Bl\ket \xi\eta~{2\o g^2}\del^n
\tr\bl[(F_{nm}+i\t{F}_{nm})\dph\br](y) 
-\xi\si_k\b{\si}_m\eta~\tr\cob_{BRST}(\b{c}D^k\dph)(y) \Br\bra \nn
&&-{\del\o\del x^0}\cob(x^0-y^0)\Bl\ket{2\o g^2}\tr\bl(\b{\psi}\b{\si}_m
\si^{0k}\eta\cob_{\xi}^{(1)}A_k\br)(y)   \Br\bra
\label{n=2WI}
\end{eqnarray}
where, for notational simplicity, we wrote the Ward-Takahashi identity
multiplied by a {\em global } $\xi$. 
We also used the equations of motion 
\begin{eqnarray}
&&D^nF_{nm}=i[D_m\dph,\phi]-i[\dph,D_m\phi]-[\b{\la},\b{\si}_m\la]
 -[\b{\psi},\b{\si}_m\psi] \\
&&\si_mD^m\la=\rt [\dph,\psi]
\end{eqnarray}
in the above derivation,  which is expected to be a safe operation. 
The term proportional to ${\del\o\del x^0}\cob(x^0-y^0)$ in (\ref{n=2WI}) can be
dropped if one uses the $\T$-product by the BJL prescription. We then
obtain 
\begin{eqnarray}
\bl[\xi Q^{(1)}(x^0),\eta S_m^{(2)}(y)\br]_{x^0=y^0}&=&
-2\rt i\Bl(\xi\eta~{2\o g^2}\del^n\tr\bl[(F_{nm}+i\t{F}_{nm})\dph\br](y) 
\nn &&
~~~~~~~~~~~~-\xi\si_k\b{\si}_m\eta~\tr\cob_{BRST}(\b{c}D^k\dph)(y)\Br)
\label{n=2com-QS}
\end{eqnarray}
after rewriting (\ref{n=2WI}) in terms of the T-product and operating the
time-derivative in the left-hand side explicitly. 
The electric charge $n_e$ and magnetic charge $n_m$ are given by
\begin{eqnarray}
&&\int d^3x ~{2\o g^2}\del_n\tr(F^{n0}\dph)=a^*\bl(n_e+{\th\o 2\pi}n_m\br)
\label{w-eff} \\
&&\int d^3x ~{2\o g^2}\del_n\tr(\t{F}^{n0}\dph)=-{4\pi\o g^2}a^*n_m
\end{eqnarray}
where $a$ is the asymptotic value of $\phi$ at spatial infinity. 
Eq.(\ref{w-eff}) represents the so called ``Witten effect'' \cite{witten}.
Note that $\th$ does  not appear in the expressions of  supercurrents,
and it enters in the superalgebra only through (\ref{w-eff}). 

The equal-time commutator of supercharges is obtained from (\ref{n=2com-QS}) by
integrating over  $d^{3}y$ as 
\begin{equation}
\bl[\xi Q^{(1)}(x^0),\eta Q^{(2)}(x^0)\br]=-2\rt i\Bl(\xi\eta Z^*
-\xi\si_k\b{\si}^0\eta\int d^3x~\tr\cob_{BRST}(\b{c}D^k\dph)(x) \Br)
\label{com-Q1Q2}
\end{equation}
where the central charge is given by
\begin{eqnarray}
&&Z=an_e+\tau an_m \\
&&\tau={\th\o 2\pi}+i{4\pi\o g^2}
\label{tau}
\end{eqnarray}
Eq.(\ref{com-Q1Q2}) has a general structure as in (\ref{uptoB-comm}), and $Z$ gives rise to the
standard formula of the central charge of $N=2$ super 
Yang-Mills theory \cite{witten-olive}. We emphasize that our derivation of
(\ref{com-Q1Q2})$\sim$ (\ref{tau}) is fully quantum mechanical in the framework of path
integral,
which is based on the Lorentz covariant Landau gauge.

\section{$D=10$ massive superparticle (D-particle)}

In this section we analyze the super algebra for a $D=10$ massive
superparticle in the BRST framework. Due to the well-known complications
arising from the off-shell non-closure of 
$\kappa$-symmetry, a straightforward prescription does not work.
Nevertheless,
we can devise a working prescription for this case also.
The Lagrangian of a massive superparticle is given by \cite{aganagic}
\begin{equation}
{\cal L}_0 ={\Pi_m^2 \over 2e}-{M^2 \over 2}e -iM
\bar{\th}\Ga_{11}\dot{\th}
\label{2nd-lag}
\end{equation} 
where $e$ is the einbein , $M$ is the mass of the superparticle and we
defined 
\begin{equation}
\Pi_m = \dot{X}_m-i\bar{\th}\Ga_m\dot{\th} 
\end{equation} 
The index $m$ runs from $0$ to $9$, and $\theta$ is a 32-component Majorana
spinor. Here $\Ga^m $ are the 10-dimensional Dirac matrices and
$\Gamma_{11}$ is a counter part of $\gamma_{5}$. We have $\bar{\theta}\eta
= \bar{\eta}\theta$
 and $\bar{\theta}\Gamma_{m}\eta = - \bar{\eta}\Gamma_{m}\theta$ for two 
Majorana spinors. 
Instead of (\ref{2nd-lag}), we here use the first order Lagrangian
\begin{equation}
{\cal L}_0=P_m\Pi_m- \frac{e}{2}(P_{m}^{2} + M^{2})-iM
\bar{\th}\Ga_{11}\dot{\th}
\label{1st-lag}
\end{equation}
where $\ps \equiv P_{m}\Gamma^{m}$.
In the chiral notation, the above Lagrangian is re-written as 
\begin{equation}
{\cal L}_0 = P_{m}\dot{X}_{m} - i\bar{\th}_L \ps \dot{\th}_L
-i\bar{\th}_R\ps \dot{\th}_R  -iM(\bar{\th}_R\dot{\th}_L
-\bar{\th}_L\dot{\th}_R ) - \frac{e}{2}(P_{m}^{2} + M^{2})
\end{equation}
The action thus defined has a local $\ka$-symmetry
\begin{eqnarray}
\cob_{\ka}\th &=& (M+ \not \!\! P \Ga_{11})\ka ,\ \ \ \cob_{\ka}P_{m} = 0
\nonum \\
\cob_{\ka}X^m &=& i\bar{\th}\Ga^m \cob_{\ka}\th \nonum \\
\cob_{\ka}e &=& 4i\dot{\bar{\th}}\Ga_{11}\ka
\end{eqnarray}  

The so-called ``irreducible $\ka$-symmetry''\cite{aganagic,berg-kallo}
 is to choose $\ka =\ka_R\equiv \frac{1}{2}(1-\Gamma_{11})\ka$.
Under this symmetry, field variables  transform as follows:
\begin{eqnarray}
\cob \th_R &=& M\ka_R , \ \ \ \cob \th_L = -\ps \ka_R\nonum \\
\cob X^m &=& i\bar{\th}_R\Ga^m M\ka_R-i\bar{\th}_L\Ga^m\ps \ka_R \nonum \\
\cob e &=& -4i\dot{\bar{\th}}_L\ka_R,\ \ \ \cob P_{m} = 0
\label{irka-sym}
\end{eqnarray}
We choose the gauge conditions $e=1$ and $\theta_{R} = 0$, which are
effected by the gauge fixing and Faddeev-Popov terms given by 
\begin{eqnarray}
{\cal L}_g 
&=&B(e-1) + ib(\dot{c}e+c\dot{e} )    +\bar{\tilde{\xi}}_L\th_R       
\label{Lg-D0} 
\end{eqnarray}
One may understand (\ref{Lg-D0}) as formally obtained from 
\begin{eqnarray}
{\cal L}_g &=& \cob_{BRST} \Bigl[b(e-1)+\bar{\bt}_L\th_R \Bigr] \nonum \\
           & =& B(e-1) + ib(\dot{c}e+c\dot{e} - 4i\dot{\bar{\theta}}_{L}
\gamma_{R})      +\bar{\xi}_L\th_R  
   -\bar{\bt}_L (ic\dot{\th}_R - M\gamma_{R})\nonumber\\
&=& B(e-1) + ib(\dot{c}e+c\dot{e})   
  +\bar{\tilde{\xi}}_L\th_R  
   + \bar{\tilde{\bt}}_L M\gamma_{R} 
\label{Lg-orig}
\end{eqnarray}
where $\gamma_{R}$ is the ghost related to the $\kappa$-symmetry, and 
$\bar{\tilde{\bt}}_L \equiv \bar{\bt}_{L} + 4b\dot{\bar{\theta}}_{L}$ and 
$\bar{\tilde{\xi}}_{L} \equiv \bar{\xi}_{L} + i
\frac{d}{dt}(c\bar{\beta}_{L})$
after partial integration.  
The variables  $c,b$ are  fermionic reparametrization ghost and anti-ghost,
respectively,  and $(\bar{\bt}_L, \bar{\xi}_{L})$ are  the
Nakanishi-Lautrup fields to 
implement $\theta_{R} = 0$. The parameter $\ep$ is a Grassmann parameter.

The ghost sector $\bar{\tilde{\bt}}_L M\gamma_{R}$ in the final expression
of (\ref{Lg-orig}) is analogous to that of the unitary gauge 
in Higgs mechanism,
and it is expected that $\tilde{\bt}_{L}$ and $\gamma_{R}$ do not
contribute  to the discontinuity in unitarity relations.   One may then
path integrate out 
$\tilde{\beta}_{L}$ and then $\gamma_{R}$
, and after the path integration  the following (nil-potent) 
reparametrization  BRST symmetry remains in the action defined by 
(\ref{Lg-D0})
\begin{eqnarray}
\cob X^m &=& -i\ep(c\dot{X}^m ),\ \ \cob e =-i\ep(\dot{c}e+c\dot{e})
 \nonum \\
\cob \th_R &=& -i\ep(c\dot{\th}_R),\ \ \cob \th_L = -i\ep(c\dot{\th}_L)
\nonum \\
\cob c &=& -i\ep c \dot{c}, \ \ \cob P_{m} = -i\ep (c\dot{P}_{m})  \nonum
\\
\cob b  &=& \ep B, \ \ \cob  B = 0 \nonum \\
\cob \bar{\tilde{\xi}}_{L} &=& -i\ep{d \over dt}(c\bar{\tilde{\xi}}_{L})
\end{eqnarray}
where the transformation property of $\tilde{\xi}_L$ is regarded to be
induced by the original transformation law
\begin{equation}
\cob \bar{\bt}_L = \ep\bar{\xi}_L, \ \ \cob \bar{\xi}_L = 0 
\end{equation}
What we have achieved in (\ref{Lg-D0}) is that the irreducible 
$\kappa$-symmetry in
(\ref{irka-sym}) is effectively gauge fixed without a ghost 
for $\kappa$-symmetry by
using 
the algebraic $\theta_{R} = 0$ gauge and thus the issue of off-shell
non-closure of $\kappa$-symmetry is avoided. 
 See also Ref.\cite{kallosh} for a recent Hamiltonian analysis of this
problem 
by using a different form of ``reparametrization'' symmetry which exists in
the superparticle.

The global target space SUSY is defined by  
\begin{eqnarray}
\cob_{SUSY} \th_{L,R} &=& \ep_{L,R} \nonum \\
\cob_{SUSY} X^m &=& i\bar{\ep}_L\Ga^m\th_L +i\bar{\ep}_R\Ga^m\th_R  \nonum
\\
\cob_{SUSY} e &=& 0,\ \ \ \cob_{SUSY}P_{m} = 0
\label{s-trf-D0}
\end{eqnarray}

The variation of the total action $S=S_0+S_g$ under SUSY transformation
with
 localized (time dependent) parameters  $\ep_{L,R}(t)$ is
\begin{equation}
\cob S =\int dt \Bigl(i\dot{\bar{\ep}}_LQ_R
+i\dot{\bar{\ep}}_RQ_L +\bar{\ep}_R\tilde{\xi}_L \Bigr)
\label{s-var-D0}
\end{equation}  
which defines supercharges
\begin{eqnarray}
&& Q_R= 2(\ps\th_L-M\th_R)   \\
&& Q_L =2(\ps\th_R+M\th_L)
\end{eqnarray}
Equations of motion for $Q_{L,R}$(or Ward-Takahashi identities) are
obtained by using  (\ref{s-var-D0}) in the path integral 
\begin{eqnarray}
{d \over dt}Q_R(t) &=& 0 \nonum \\
{d \over dt}Q_L(t) &=& -i \tilde{\xi}_L(t) 
\label{eq-QL}
\end{eqnarray}

The charge $Q_L$ is not conserved due to the term
$\bar{\tilde{\xi}}_L\th_R$ in
${\cal L}_g$. However one can find the supercharge which is conserved
up to a BRST exact term. For this purpose, we rewrite the equation 
(\ref{eq-QL}). 
From the equation of motion for $\theta_{R}$ we obtain 
\begin{equation}
\frac{\partial S_{0}}{\partial \theta_{R}} + \frac{\partial S_{g}}{\partial
\theta_{R}} = \frac{\partial S_{0}}{\partial \theta_{R}} +
\bar{\tilde{\xi}}_{L}  = 0
\label{eq-thR}
\end{equation}
where we wrote the total action as a sum of the main part $S_{0}$ and the 
gauge fixing part $S_{g}$. 

On the other hand , the $\kappa$-symmetry of $S_{0}$ suggests
\begin{equation}
\int dt\lf[ \frac{\partial S_{0}}{\partial \theta_{R}(t)}M\kappa_{R}(t) 
+ \frac{\partial S_{0}}{\partial \theta_{L}(t)}\delta\theta_{L}(t)
+ \frac{\partial S_{0}}{\partial X^{m}(t)}\delta X^{m}(t)
+ \frac{\partial S_{0}}{\partial e(t)}\delta e(t)\ri] = 0
\end{equation}
Since $X^{m}$ and $\theta_{L}$ do not appear in $S_{g}$, we have the
equations of motion
\begin{equation}
\frac{\partial S_{0}}{\partial \theta_{L}(t)} = \frac{\partial
S_{0}}{\partial X^{m}(t)} = 0
\end{equation}
The equation of motion for $e(t)$ is 
\begin{equation}
\frac{\partial S_{0}}{\partial e(t)} + \frac{\partial S_{g}}{\partial e(t)}
= 0
\end{equation} 
We thus obtain
\begin{eqnarray}
\int dt \frac{\partial S_{0}}{\partial \theta_{R}(t)}M\kappa_{R}(t)
&=& \int dt \frac{\partial S_{g}}{\partial e(t)}\delta e(t) \nonumber\\
&=& \int dt \bl(B + ib\dot{c} - 
i\frac{d}{dt}(bc)\br)(- 4i\dot{\bar{\theta}}_{L}\kappa_{R})
\label{ka-of-S}
\end{eqnarray}
If we combine the above equations (\ref{eq-thR}) and 
(\ref{ka-of-S}), we obtain
\begin{equation}
M\bar{\tilde{\xi}}_{L} = 4 \frac{d}{dt}(bc\dot{\bar{\theta}}_{L})
  +4i\bl( B\dot{\bar{\theta}}_{L} + b\frac{d}{dt}(ic\dot{\bar{\theta}}_L)
\br)
\end{equation}
Thus we finally  obtain
\begin{eqnarray}
\frac{d}{dt}Q_{L} &=& -i \bar{\tilde{\xi}}_{L} \nonumber\\
&=& -\frac{4i}{M}\frac{d}{dt}(bc\dot{\bar{\theta}}_L) 
    +\frac{4}{M}\delta_{BRST} (b\dot{\bar{\theta}}_{L})
\end{eqnarray}
We may thus define
\begin{equation}
\tilde{Q}_{L}(t) = Q_{L}(t)+q_L(t) \equiv
Q_{L}(t) +\frac{4i}{M}b(t)c(t)\dot{\theta}_{L}(t)
\end{equation}
which satisfies
\begin{equation}
\frac{d}{dt}\tilde{Q}_{L} =  \frac{4}{M}
\delta_{BRST}(b\dot{\theta}_{L})
\end{equation}
namely, $\tilde{Q}_{L}$ is conserved up to a {\em BRST exact term}.
A relation of the structure of this last equation is also  derived for 
a  supermembrane\cite{berg} , for example,  by observing that the
$\kappa$-symmetry
transformation does not interfere with the reparametrization symmetry
properties\cite{fuji-oku}.

Note that $q_{L}$ is  {\em not} BRST closed $\cob_{BRST}(q_{L}) \not = 0$.
Therefore $q_{L}$ cannot be written in  a BRST exact form  $q_{L}\not =
\cob_{BRST}(\ast)$.
In the context of our general analysis, the gauge fixing fermion $\Psi$ in
the present case is regarded to be given by (see (\ref{Lg-orig}))
\begin{equation}
\Psi \equiv b(e-1) + \bar{\beta}_{L}\theta_{R}
\end{equation}
One can then confirm  for a localized $\delta_{SUSY}$
\begin{eqnarray}
\delta_{SUSY}\delta_{BRST}\Psi &=& \epsilon \bar{\xi}_{L}\epsilon_{R}(t)
- i\bar{\beta}_{L}c\dot{\epsilon}_{R}(t)\nonumber\\
&\neq& \epsilon \bar{\xi}_{L}\epsilon_{R}(t)\nonumber\\
&=&  \delta_{BRST}\delta_{SUSY}\Psi
\end{eqnarray}
Therefore the relation (\ref{mod-charge}) does not hold in the present
example, namely,
$\cob_{BRST}(q_{L}) \not = 0$. However, one can confirm that
\begin{equation}
\delta^{\prime}_{SUSY}\delta_{SUSY}\delta_{BRST}\Psi =
\delta_{BRST}\delta^{\prime}_{SUSY}\delta_{SUSY}\Psi = 0
\label{SSB-D0}
\end{equation}
( in fact, $\delta^{\prime}_{SUSY}\delta_{SUSY}\delta_{BRST}( e,
\theta_{R})
=
\delta_{BRST}\delta^{\prime}_{SUSY}\delta_{SUSY}(e,\theta_{R}) = 0$) and
thus the
property (\ref{uptoB-comm})
is expected to hold. This is confirmed by the following explicit
calculation. 

 It is important to observe that the extra term  $q_L(t)$ in
 $\tilde{Q}_{L}(t)$
contains the derivative of $\theta_{L}$ variable. This means that the 
localized  supersymmetry transformation of $q_L(t)$
 gives rise to a term proportional to $\partial_{t}\delta(t - t^{\prime})$,
which is eliminated by BJL prescription.  
Consequently, it does not influence the equal-time SUSY algebra (i.e., 
$Q_{L}$ and $\tilde{Q_{L}}$ satisfy the same form of algebra ).
Therefore,we analyse the anti-commutator of $Q_{L,R}$ in the rest of
this section.

If we make the change of variables generated by a {\em localized} SUSY 
transformation (\ref{s-trf-D0}) in  the path integral defined by, 
for example, 
\begin{equation}
\langle Q_{R\beta}(s)\rangle \equiv \frac{1}{Z}\int d\mu
Q_{R\beta}(s)
e^{iS}
\end{equation}
where $Z=\int d\mu e^{iS}$,
we obtain the following relations (Ward-Takahashi identities),
\begin{eqnarray}
  {d\over dt}\Bigl\ket\Ts Q_R^{\al}(t)Q_{R\bt}(s)\Bigr\bra
&=&\lf\ket2\ps ^{\al}_{~\bt}\cob(t-s)\ri\bra      \nonum \\
  {d\over dt}\Bigl\ket\Ts Q_L^{\al}(t)Q_{L\bt}(s)\Bigr\bra
&=&\lf\ket2\ps ^{\al}_{~\bt}\cob(t-s)-\Ts
i\tilde{\xi}_L^{\al}(t)Q_{L\bt}(s) \ri\bra      \nonum \\
  {d\over dt}\Bigl\ket\Ts Q_L^{\al}(t)Q_{R\bt}(s)\Bigr\bra
&=&\lf\ket2M\cob^{\al}_{~\bt}\cob(t-s)-\Ts
i\tilde{\xi}_L^{\al}(t)Q_{R\bt}(s) \ri\bra      \nonum \\
  {d\over dt}\Bigl\ket\Ts Q_R^{\al}(t)Q_{R\bt}(s)\Bigr\bra
&=&\lf\ket-2M\cob ^{\al}_{~\bt}\cob(t-s)\ri\bra 
\label{Ts-D0}     
\end{eqnarray}
All the operator ordering is defined by $\Ts$-product in path
integral.
We have no terms which contain  $\del_t \cob(t-s)$, and we obtain the
$\T$-product relations by using the BJL prescription
\begin{eqnarray}
  {d\over dt}\Bigl\ket\T Q_R^{\al}(t)Q_{R\bt}(s)\Bigr\bra
&=&\lf\ket2\ps ^{\al}_{~\bt}\cob(t-s)\ri\bra      \nonum \\
  {d\over dt}\Bigl\ket\T Q_L^{\al}(t)Q_{L\bt}(s)\Bigr\bra
&=&\lf\ket2\ps ^{\al}_{~\bt}\cob(t-s)-\T
i\tilde{\xi}_L^{\al}(t)Q_{L\bt}(s) \ri\bra      \nonum \\
  {d\over dt}\Bigl\ket\T Q_L^{\al}(t)Q_{R\bt}(s)\Bigr\bra
&=&\lf\ket2M\cob^{\al}_{~\bt}\cob(t-s)-\T
i\tilde{\xi}_L^{\al}(t)Q_{R\bt}(s) \ri\bra      \nonum \\
  {d\over dt}\Bigl\ket\T Q_R^{\al}(t)Q_{R\bt}(s)\Bigr\bra
&=&\lf\ket-2M\cob ^{\al}_{~\bt}\cob(t-s)\ri\bra   
\label{D0-WI} 
\end{eqnarray}
If one uses the Lagrangian (\ref{2nd-lag}) instead of (\ref{1st-lag}), the
$\Ts$-product 
gives rise to identities with extra terms proportional to  $\del_t
\cob(t-s)$.

When one explicitly operates the time derivative in the left-hand sides of
(\ref{D0-WI}), one obtains equal-time anti-commutator by using the
equations of
motion in (\ref{eq-QL}). Comparing the both sides of (\ref{D0-WI}), 
one thus obtains  
\begin{eqnarray}
\{Q_R^{\al}(t),Q_{R\bt}(t)\} &=& 2\ps ^{\al}_{~\bt} \nonum \\
\{Q_L^{\al}(t),Q_{L\bt}(t)\} &=& 2\ps ^{\al}_{~\bt} \nonum \\
\{Q_L^{\al}(t),Q_{R\bt}(t)\} &=& 2M\cob^{\al}_{~\bt} \nonum \\
\{Q_R^{\al}(t),Q_{L\bt}(t)\} &=& -2M\cob^{\al}_{~\bt}
\label{s-alg-D0}
\end{eqnarray}
where  $M$ can be thought of as a ``central charge''. These relations
reproduce  the expected  form (written in a chiral notation) of
superalgebra 
\begin{equation}
[\bar{\ep}Q,\bar{\ep}'Q]= 2\bar{\ep}(\ps+M\Ga_{11})\ep' 
\end{equation}  
Note that we do not obtain the BRST exact pieces in the right-hand sides of
(\ref{s-alg-D0}) due to (\ref{SSB-D0}). 

One can confirm that all the correlation functions in (\ref{Ts-D0}) are
well-defined .
The essence of our path integral is described by
the Lagrangian
\begin{equation}
\tilde{{\cal L}} \equiv P_{m}\dot{X}^{m} -i\bar{\theta}_{L}\ps
\dot{\theta}_{L}
-i\bar{\theta}_{R}\ps \dot{\theta}_{R} -iM(\bar{\theta}_{R}\dot{\theta}_{L}
-
\bar{\theta}_{L}\dot{\theta}_{R}) -\frac{1}{2}( P_{m}^{2} + M^{2}) +
\bar{\tilde{\xi}}_{L}\theta_{R} 
\label{D0-L'}
\end{equation}
after path integration over other unphysical variables. 
The path integral is then defined by , for example, 
\begin{equation}
\langle \Ts Q_{R}(t)Q_{R}(s)\rangle = \int {\cal D}P_{m}
{\cal D}X^{m}{\cal D}\theta_{L}{\cal D}\theta_{R}{\cal D}\tilde{\xi}_L
Q_{R}(t)Q_{R}(s) e^{i\int \tilde{{\cal L}} dt }/ \int d\mu
e^{i\int \tilde{{\cal L}} dt }
\end{equation}
The extra factor arising from the second-class constraint\cite{kallosh},
which may be represented by introducing an extra {\em bosonic} spinor
$\phi_{L}$ described by ${\cal L}^{\prime} = -i\bar{\phi}_L \!\!\ps
\phi_{L}$,
does not influence our correlation functions.
It is significant that the Hamiltonian is given by $H =
\frac{1}{2}(P_{m}^{2}
+ M^{2}) - \bar{\tilde{\xi}}_{L}\theta_{R}$ in (\ref{D0-L'}), and the
quantum 
fluctuation of 
$ \bar{\tilde{\xi}}_{L}\theta_{R}$ term generally smears the pole 
position of the 
Hamiltonian  $H = \frac{1}{2}(P_{m}^{2} + M^{2})$ . We can thus avoid the
appearance of singular correlation functions in (\ref{Ts-D0}). The
advantage of
the present  path integral approach is that we can directly obtain the
physical results  without going through the Dirac bracket manipulation. 

We emphasize that  our treatment of $\kappa$-symmetry  also  works for a 
supermembrane\cite{berg}\cite{fuji-oku}, for example.

\section{Discussion}

We have illustrated the simple rules in (\ref{mod-charge}) - (\ref{uptoB-comm}) for a general class
of 
supersymmetric gauge theories when quantized by the BRST procedure. In our 
formulation, the Faddeev-Popov ghosts and Nakanishi-Lautrup fields are 
all assigned to be SUSY scalar. In practical applications, our method 
directly gives rise to desired results without going through the (often
complicated) Dirac bracket analysis. Our treatment of these algebraic 
relations are in the full quantum mechanical level up to possible quantum
anomalies. 

It is possible to incorporate the conventional gauge invariant ( but not
necessarily supersymmetric) regularization into our scheme. We can thus
readily 
derive the anomalous commutator such as ( in the one-loop level accuracy)
\begin{equation}
[\xi Q, R^{0}(x)] = - \xi S^{0}(x) + i \frac{Ng^{2}}{4\pi^{2}} \tr
(\bar{\lambda}\bar{\sigma}^{lm}F_{lm}\bar{\sigma}^{0}\xi)
\end{equation}
for the supercharge $Q_{\alpha}$ and the chiral $R$ - symmetry current
$R^{m}(x) = 2 \tr (\bar{\lambda}\bar{\sigma}^{m}\lambda)$ for the Lagrangian
(\ref{n=1Lag}), by using the conventional chiral anomaly for 
$\partial_{m}R^{m}(x)$
and the BJL prescription. However, it is not easy to make a global 
supersymmetric structure manifest in our component formulation in the Wess-
Zumino gauge. As a result, we have not succeeded so far in identifying the 
possible central extension of $N=1$ SUSY algebra suggested in
Ref.\cite{shifman}
 in a reliable way; the suggested possible central extension is related to 
quantum anomaly and written as an integral of total (spatial) divergence.
 What we need is a generalization of the path integral evaluation in
Ref.\cite{konishi} and a supersymmetric generalization of the analyses in
Ref.\cite{fuji}, for example.

With this limitation in  mind, we still believe that our simple rules in
(\ref{mod-charge})  - (\ref{uptoB-comm}) should be useful in the practical 
analyses of SUSY algebra
and its
physical implications.
 

\end{document}